\documentclass{article}
\usepackage{arxiv}

\usepackage[utf8]{inputenc} 
\usepackage[T1]{fontenc}    
\usepackage[hidelinks]{hyperref}       
\usepackage{url}            
\usepackage{booktabs}       
\usepackage[version=4]{mhchem}
\usepackage{amsfonts}       
\usepackage{mathtools}
\usepackage{nicefrac}       
\usepackage{lipsum}		
\usepackage{graphicx}
\usepackage[square,numbers]{natbib}
\usepackage{caption}
\usepackage{subcaption}
\usepackage{amsmath}

\usepackage{booktabs}

\usepackage{bm}

\usepackage{url}
\usepackage{siunitx}
\DeclareSIUnit\atmosphere{atm}
\usepackage{pifont}

\usepackage[dvipsnames]{xcolor}
\definecolor{newcolor}{rgb}{.8,.349,.1}
\graphicspath{{Figure/}}

   \newcommand{\mean}[1]{\overline{#1}\,}

   \newcommand{\logmean}[1]{\overline{#1}^{\text{log}}}
   
   \newcommand{\gmean}[1]{\overline{#1}^{G}}
   \newcommand{\hmean}[1]{\overline{#1}^{H}}

  \newcommand{\meantwo}[1]{\overline{#1}^{\,\#}}
  \newcommand{\meanone}[1]{\overline{#1}^{\,\star}}
  \newcommand{\meanless}[1]{\overline{#1}^{\,<}}
  \newcommand{\meanmore}[1]{\overline{#1}^{\,>}}
   \renewcommand{\d}{\mathrm{d}}

\newcommand{\bff}{\mathbf{f}}

\newcommand{\dtp}{\delta^{\,+}}

\DeclareMathOperator{\sgn}{sgn}

\usepackage{amssymb}
\usepackage{amsthm}
\theoremstyle{remark}
\newtheorem{remark}{Remark}

   \usepackage{caption}
   \usepackage{subcaption}
   \usepackage[normalem]{ulem}
\usepackage{upgreek}
\title{Entropy-stable discretizations for the compressible Euler equations using simple adaptive averages\thanks{Distribution Statement A: Approved for Public Release; Distribution is Unlimited. AFRL-2026-2473.}}%

\date{May 19, 2026}	

\author{
    \href{https://orcid.org/0000-0002-6518-3114}{\includegraphics[scale=0.06]{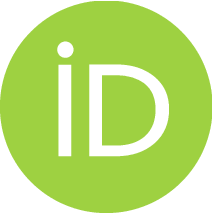}\hspace{1mm} Carlo {De~Michele}}\\
	Gran Sasso Science Institute (GSSI)\\
	L'Aquila, Italy \\
	\texttt{carlo.demichele@gssi.it} \\
    \And
    \href{https://orcid.org/0000-0002-7463-559X}{\includegraphics[scale=0.06]{orcid.eps}\hspace{1mm}Ayaboe K.~Edoh} \\
    Amentum~--- Edwards Air Force Base\\
    CA 93524, USA\\
    \texttt{ayaboe.edoh@us.amentum.com}\\
}

\renewcommand{\shorttitle}{ES discretizations for the compressible Euler equations using simple adaptive averages}%

\hypersetup{
pdftitle={Entropy-stable discretizations for the compressible Euler equations using simple adaptive averages},
pdfsubject={physics.flu-dyn},
pdfkeywords={Compressible flows, Entropy stability,Kinetic-energy preserving, Pressure-equilibrium-preserving},
pdfauthor={Carlo De~Michele, Ayaboe K.~Edoh}
}
\begin{document}
\maketitle

\begin{abstract}
Entropy stabilization of the compressible Euler system is achieved by adapting the averages that are applied to the density and internal energy variables. 
The approach achieves non-linear robustness despite the use of simplified symmetric means (e.g., arithmetic, geometric, or harmonic evaluations), including their related expansions for asymptotic entropy conservation.
The proposed formulation works via centralized convective terms and can naturally adhere to additional structures of the flow equations such as kinetic-energy- and pressure-equilibrium-preservation.
\end{abstract}


\section{Introduction}
Numerical simulations of compressible flows---especially in under-resolved turbulent or transitional regimes---remain a challenge in computational fluid dynamics. In this context, discretizations that respect the physical principle of an entropy inequality, consistent with the second law of thermodynamics, have been of great interest for ensuring non-linear robustness and physical fidelity \cite{Gassner_JCP_2016,Rojas_JCP_2021,Aiello_arXiv_FTaC_2026}.

Traditionally, entropy stabilization is achieved by appending tailored dissipative terms to the baseline semi-discretization \cite{Edoh2024,Chan2025,Bercik2026}. 
Instead, the present work proposes a novel framework for achieving entropy stability (ES) by locally adapting the central averages employed within the inviscid fluxes of the compressible Euler equations.
Isolating such choices to the density and internal energy interface reconstructions allows to identify simple ES conditions which rely on employing averages (e.g., arithmetic, geometric, harmonic) that are strictly above or below the logarithmic means typically used to recover EC schemes \cite{Ismail_JCP_2009,Ranocha_JSC_2018,DeMichele_JCP_2025}.
The resulting methodology is comprised of central flux averages of the inviscid terms, which suggests a dispersive-type mechanism within the primary equations, rather than a dissipative one. 
The formulation can furthermore be made to satisfy critical additional structures such as kinetic-energy-preservation (KEP) \cite{Jameson_JSC_2008b,Veldman_JCP_2019,Coppola_JCP_2019,Edoh_JCP_2022} and pressure-equilibrium-preservation (PEP) \cite{Shima_JCP_2021,DeMichele_JCP_2024}, properties typically not preserved by traditional ES dissipation.
The new concept is leveraged herein to develop entropy-stable extensions for the class of asymptotically entropy-conservative (AEC) schemes \cite{DeMichele_JCP_2023,DeMichele_JCP_2024}.
Also a regularization strategy for sharp gradients (e.g., shocks) is presented. 

\section{Problem Formulation}

The current work considers the compressible Euler equations written in a semi-discrete flux conservative form.
For simplicity of the present development, we consider a uniform grid in one dimension with spacing $h$ and locations $x_i$, along with second-order two-point fluxes. 
Extension to mult-dimensions is then straightforward by tensor products. 
For a generic quantity $\phi$, we define the two-point backward-difference operator as $\delta^- \phi = (\phi_{i}- \phi_{i-1})$ and the forward-difference operator as $\delta^+ \phi = (\phi_{i+1}- \phi_{i})$. 
We furthermore identify the arithmetic mean $\overline{\phi} = (\phi_{i+1} + \phi_i)/2$, the geometric mean $\gmean{\phi} = \sqrt{\phi_{i+1}\phi_i}$ and the harmonic mean $\hmean{\phi} = \left(\gmean{\phi}\right)^2/\mean{\phi}$.

The form of the baseline conservative one-dimensional semi-discretization assumed herein is
\begin{equation}\label{eq:discretization}
    \partial_t {\bf q} + \partial_x  \bff({\bf q}) 
 = 0 \ \to \ \frac{\mathrm{d}}{\mathrm{d}t}  \left[\begin{array}{c} \rho \\ \rho u \\ \rho E \end{array} \right]_i + \frac{1}{h} \cdot \delta^- \left[\begin{array}{l} \mathcal{F}_\rho \\ \mathcal{F}_{\rho u} \\ \mathcal{F}_{\rho E} \end{array} \right]_i = 0,  \quad \text{with} \
\left\{ \begin{array}{r l}
\mathcal{F}_\rho|_i =& \meanone{\rho} \overline{u} \\
\mathcal{F}_{\rho u}|_i =& \mathcal{F}_\rho|_i \cdot \overline{u} + \overline{p} \\
\mathcal{F}_{\rho E}|_i =& \mathcal{F}_\rho|_i \cdot \left(   \frac{u_i u_{i+1}}{2} + \meantwo{e} \right)\\
& \qquad+ \frac{1}{2} \left(p_iu_{i+1} + p_{i+1}u_i \right)
 \end{array} \right.
\end{equation}
The above corresponds to the conservative variables for a single-component gas, ${\bf q} = [\rho, \rho u, \rho E]^\text{T}$, whose temporal evolution is driven by derivatives of the associated fluxes, $\mathcal{F}$. 
These are written in terms of the density $\rho$, the velocity $u$, the pressure $p$, and the total energy per unit mass $E$, which is composed of kinetic and internal energy, $E = (e+\kappa)$ with $\kappa = u^2/2$.
Assuming a \emph{calorically perfect} gas, the internal energy and pressure are related to the temperature $T$ as
$e = e_\text{ref} +c_v T$, $p = \rho R T = (\gamma -1) \cdot \rho e$, where $c_v$ and $c_p$ are the constant specific heats at constant volume and pressure, $\gamma = c_p/c_v$ is the specific heat ratio, and $c_p - c_v = R$ with $R$ being the gas constant.
In Eq.~\eqref{eq:discretization}, the convective components of the inviscid fluxes are $\mathcal{F}|_i^{c} = [\mathcal{F}_\rho,\mathcal{F}_{\rho u}, \mathcal{F}_{\rho E = \rho (e + \kappa)} ]|_i^c = (\meanone{\rho} \overline{u}) \cdot [1, \mean{u},(\meantwo{e} + \gmean{\kappa})]$, where $\meanone{\rho}$ and $\meantwo{e}$ remain to be specified.

The flux schemes considered in the present work are assumed to satisfy the KEP structure with respect to the convective part of the momentum flux \cite{Jameson_JSC_2008} in addition to the notion of energy consistency \cite{Kuya_JCP_2018,DeMichele_C&F_2023}, such that the total energy flux is constructed to be compatible with the implied discrete kinetic energy equation plus the internal energy evolution. 
Working within this specific family of KEP and energy-consistent flux averages, the convective fluxes for mass and internal energy still remain free to be defined with respect to $\meanone{\rho}$ and $\meantwo{e}$.
This offers the possibility of imposing additional structural properties such as pressure-equilibrium preservation.

The present work aims to construct a class of schemes that is not only KEP and PEP but also \textit{entropy stable (ES)} for a calorically perfect gas via the judicious selection of $\meanone{\rho}$ and $\meantwo{e}$, in the sense that the method satisfies a discrete form of the physical entropy inequality upon integration over the domain,
\begin{equation}\label{eq:entropy_stability}
\frac{\mathrm{d}}{\mathrm{d}t} \int_\Omega \rho s \geq 0 \quad \text{with} \quad s - s_\text{ref} = c_v \log T - R \log \rho \overset{\text{(CPG)}}=  c_v\log{e} - R\log{\rho} \ .
\end{equation}

\section{Entropy error analysis}

A discrete form of the entropy integration of Eq.~\eqref{eq:entropy_stability} is given by
\begin{equation}
 h \cdot \sum_i  \left(\frac{\mathrm{d}}{\mathrm{d}t}\  \rho s\right)_i 
  =  \overbrace{-\sum_i\left[ \left(c_v \log e - R\log \rho - c_p\right) \cdot \delta^- (\meanone{\rho}\overline{u}) +  \frac{c_v}{e} \cdot  \delta^- \left(\meanone{\rho }\meantwo{e}\overline{u}\right) + \frac{p}{T} \cdot  \delta^- \overline{u} \right]_i }^{ \sum_i \left[ \left(s - c_p\right) \cdot \delta^- \mathcal{F}_\rho \ +\   \frac{1}{T} \cdot \left(\delta^-\mathcal{F}_{ \rho e}^c + p \cdot \delta^- \overline{u} \right)\right]_i} 
\end{equation}
where the above leverages the fact that the discretization is KEP as well as energy consistent in both its convective and pressure contributions (see also \cite{DeMichele_JCP_2025} for details). 
Applying the summation-by-parts identity and ignoring boundary terms (i.e., $a \cdot \delta^- b = - b \cdot \delta^+ a$) 
gives the following sufficient point-wise conditions for entropy stability:
\begin{eqnarray}\label{eq:ES_conditions}
 \left[ R\,\overline{u}\cdot \left(   \delta^+ \rho - \meanone{\rho} \cdot \delta^+ \log \rho \right)  + c_v\, \meanone{\rho}   \overline{u} \cdot  \left( \meantwo{e} \cdot \delta^+ e^{-1} - \delta^+ \log e^{-1} \right)\right]_i \ge 0 
\overset{\text{implies}}\impliedby   \left\{\begin{array}{l}    
\left(\mean{u} \, \left(\dtp \log{\rho}\right)_i\right) \cdot \left[ \logmean{\rho} - \meanone{\rho} \right]\geq0 \\ \\ \left(- \overline{u} \, \left(\delta^+ e^{-1}\right)_i\right) \cdot \left[\overline{e}^{H\log} - \meantwo{e}\right] \ge 0 
\end{array}
\right.
\end{eqnarray}
where $\logmean{\rho} \triangleq \delta^+ \rho/\delta^+ \log \rho$ and $\overline{e}^{H\log} \triangleq \delta^+ \log e^{-1}/\delta^+ e^{-1}$ are the definitions of the logarithmic means. 
Entropy conservation is thus achieved by setting $\meanone{\rho} = \logmean{\rho}$ and $\meantwo{e} = \overline{e}^{H\log}$, which would then have Eq.~\eqref{eq:discretization} correspond to the EC scheme of Ranocha~\cite{Ranocha_CAMC_2021}.
However, choosing to have each term in Eq.~\eqref{eq:ES_conditions} be positive recovers an ES scheme. 
An ES flux can thus be obtained through an appropriate local choice of the averages used in the numerical fluxes for density and internal energy; depending on the sign of the product between the velocity and the thermodynamic gradients, one needs to use a mean that is either larger $\meanmore{(\cdot )}$ or smaller $\meanless{(\cdot )}$ than the one enforcing EC.

    \begin{table}[h!]
\centering
\renewcommand{\arraystretch}{1.3}
\setlength{\tabcolsep}{10pt}
\begin{tabular}{c|c|c}
    & $\displaystyle \sgn(\mean{u}\dtp\! e) > 0$ 
    & $\displaystyle \sgn(\mean{u}\dtp\! e) < 0$ \\[2pt]
\hline
$\displaystyle \sgn(\mean{u}\dtp\! \rho) > 0$
&
$\begin{aligned}
\meanone{\rho} &= \meanless{\rho} ,\\
\meantwo{e}    &= \meanless{e}
\end{aligned}$
&
$\begin{aligned}
\meanone{\rho} &= \meanless{\rho} ,\\
\meantwo{e}    &= \meanmore{e}
\end{aligned}$
\\[6pt]
\hline
$\displaystyle \sgn(\mean{u}\dtp\! \rho) < 0$
&
$\begin{aligned}
\meanone{\rho} &= \meanmore{\rho} ,\\
\meantwo{e}    &= \meanless{e}
\end{aligned}$
&
$\begin{aligned}
\meanone{\rho} &= \meanmore{\rho} ,\\
\meantwo{e}    &= \meanmore{e}
\end{aligned}$
\end{tabular}
\caption{Local choices of interpolating means for density and internal energy ensuring an ES discretization. The $\meanless{\phi}$ and $\meanmore{\phi}$ indicate means respectively smaller and larger than those required to obtain an EC scheme, i.e.,~$\logmean{\rho}$ and $(\logmean{1/e})^{-1}$.}\label{tab:ES_conditions}
\end{table}

Table~\ref{tab:ES_conditions} summarizes a selection guide for the entropy-stabilizing averages, furthermore leveraging the fact that $
    \sgn(\mean{u}\dtp \log \rho) = \sgn(\mean{u}\dtp \rho)$ and $ 
    \sgn\left(-\overline{u}\, \delta^+ e^{-1}\right) = \sgn(\mean{u}\dtp e)$ in order to evaluate any additional jumps in terms of the native set $\rho$ and $e$. 
The present work leverages the well-known chain of mean inequalities~\cite{DeMichele_JCP_2023} 
$\hmean{\phi}\leq \overline{\phi}^{H\log} \leq
    \gmean{\phi}\leq\logmean{\phi}\leq\mean{\phi}$ as a starting point for proposing simple adaptive density and internal energy averages for entropy stability.
Specifically, we consider including modifying functions $\mathcal{M}$ that act on the classical means according to a two-point jump variable, such as $\langle \phi \rangle = (\delta \phi/2\overline{\phi})$.
These functions are consistent with unity (i.e., $\lim_{\langle \phi \rangle \to 0}\mathcal{M} \to 1$) and are formulated such as to remain strictly above or below the target logarithmic mean.
Here we consider the following configuration:
\begin{equation}\label{eq:modifying_functions}
\begin{array}{l l }
\meanless{\rho} = \hmean{\rho} \cdot \mathcal{M}^{<\log}_H \ \left(\text{or} \ \  \gmean{\rho} \cdot \mathcal{M}^{<\log }_G\right) \qquad
& \meanmore{\rho} =  \mean{\rho} \cdot \mathcal{M}^{>\log}_A \\
\meanless{e} =  \hmean{e} \cdot \mathcal{M}^{<H\log}_H  &
\meanmore{e} =  \mean{e} \cdot \mathcal{M}^{>H\log}_A \ \left(\text{or}  \ \ \gmean{e} \cdot \mathcal{M}^{>H\log }_G \right) \ .
\end{array}
\end{equation}
These methods may furthermore be made to maintain pressure equilibrium by enforcing
\begin{equation}\label{eq:PEP_requirements_modifying}
  \text{(PEP requirements)} \quad  \mathcal{M}^{>\log }_A \cdot \mathcal{M}^{<H\log }_H = 1 \quad \mathcal{M}^{>H\log }_A \cdot \mathcal{M}^{<\log }_H = 1 \quad 
    \mathcal{M}^{>H\log }_G \cdot  \mathcal{M}^{<\log }_G = 1 \ .
\end{equation}
In other words, achieving PEP requires the combinations ($\meanless{\rho}$, $\meanmore{e}$) and ($\meanmore{\rho}$, $\meanless{e}$) to be compatible such that $\meanone{\rho} \meantwo{\rho^{-1}} = 1$ \cite{DeMichele_JCP_2024}, in the PEP configuration of constant velocity and pressure, with other combinations of the variables not corresponding to a PEP discretization\footnote{The combinations of  ($\meanless{\rho}$, $\meanless{e}$) and ($\meanmore{\rho}$, $\meanmore{e}$) do not occur when assuming PEP because
 $\delta^+ \rho$ and $\delta^+ e$ cannot have the same sign.}. 

The proposed adaptation process modifies the central averages employed within the convective terms and thus can be understood as achieving entropy stability through dispersive-like mechanisms.
Another interpretation is that the modifying function works to bias the reconstruction such that $\meanless{\phi}$ or $\meanmore{\phi}$ moves within the mean's interval $[\phi_i,\phi_{i+1}]$, assuming that $\mathcal{M}$ is appropriately bounded.
The present dynamic procedure for structure-preservation fits in with other proposed methods that seek to recover entropy stability by leveraging otherwise non-EC discretizations \cite{Coppola_JCP_2019,doehring2026}.
Also, the dependence on thermodynamic jumps and the adjective velocity bares resemblance to the ES scheme of Eq.~(146) in \cite{Artiano2026}. 
And while this work focuses on the Euler equations, the proposed concept is straightforwardly extendable to other equations\footnote{The ES condition for the Burgers flux $f(u) = \frac{1}{2}u^2$ would be 
$(\meanone{u^2} - \overline{u^2}|_\text{EC} ) \cdot  \delta^+ u \ge 0 $, where $\overline{u^2}|_\text{EC} \triangleq \frac{2}{3} \mean{u^2} + \frac{1}{3} \gmean{u^2}$ is the Heronian average yielding entropy conservation \cite{Fisher_JCP_2013b}. For entropy stability, one can thus adapt between simple means (e.g., arithmetic and geometric) for $\meanone{u^2}$ based on the sign of the scalar's gradient.}.

\subsection{Entropy stable implementation of AEC schemes}

This section applies this proposed ES framework to the class of AEC expansions, noting and leveraging their one-sided approximation to the logarithmic means while identifying appropriate pairings for also maintaining PEP.
AEC schemes offer a practical way of approximating the logarithmic means by enhancing classical averages with a polynomial series written in terms of the jump variable $\langle \phi \rangle$,
\begin{equation}
    \begin{array}{l l l}
\mean{\phi}^{>\log}_{A,AEC(N)} = \mean{\phi} \cdot \mathcal{P}_{\alpha,N}^{-1} &
\mean{\phi}^{<\log}_{H,AEC(N)} = \hmean{\phi}  \cdot (\mathcal{P}_{\beta,N}/\mathcal{P}_{\alpha,N} 
) &
\mean{\phi}^{<\log}_{G,AEC(N)} = \gmean{\phi}  \cdot \mathcal{P}_{\gamma,N}^{-1} \\ \\
\mean{\phi}^{>H\log}_{A,AEC(N)} = \mean{\phi} \cdot (\mathcal{P}_{\beta,N}/\mathcal{P}_{\alpha,N} 
)^{-1} &
\mean{\phi}^{<H\log}_{H,AEC(N)} = \hmean{\phi}  \cdot \mathcal{P}_{\alpha,N} &
\mean{\phi}^{>H\log}_{G,AEC(N)} = \gmean{\phi}  \cdot \mathcal{P}_{\gamma,N}
    \end{array}    
\end{equation}
where we identify the following associated polynomial functions:
\begin{equation}\label{eq:polynomial_functions_AEC}
    \begin{array}{l l l}
    \mathcal{P}_{\alpha,N} = \sum_{n=0}^N \frac{\langle\phi\rangle^{2n}}{2n+1}, & 
    \mathcal{P}_{\beta,N} = \sum_{n=0}^N \langle{\phi}\rangle^{2n} & 
    \mathcal{P}_{\gamma,N} =
    \left[\sum_{n=0}^N \left(\sum_{m=0}^n (-1)^m {1/2 \choose m} \cdot \frac{1}{2(n-m)+1}\right) \langle{\phi}\rangle^{2n}\right]. 
    \end{array}
\end{equation}
In this way, the $\mathcal{P}$ functions are specific renditions of the modifying functions $\mathcal{M}$ previously mentioned in Eq.~\eqref{eq:modifying_functions}.
Such AEC schemes (also referred to as KEEP(N) methods)---of which various have been proposed \cite{DeMichele_JCP_2023,DeMichele_JCP_2024,Tamaki_JCP_2022,Kawai_JCP_2025}---approach the logarithmic means as $N\to \infty$. 
However a nominal application of such schemes for finite expansions does not guarantee strict entropy conservation or stability. 
This can be addressed by employing the adaptive ES procedure proposed in the previous section, which would leverage the fact that the AEC schemes strictly remain on one side of the target logarithmic mean.
This observation stems from the fact that the polynomial functions of Eq.~\eqref{eq:polynomial_functions_AEC} are comprised of even powers with semi-definite coefficients; 
combined with the established convergence as $N \to \infty$, then this establishes the one-sided requirement for the AEC schemes\footnote{For the rational function $\mathcal{P}_{\alpha \beta,N} \triangleq (\mathcal{P}_{\beta,N}/\mathcal{P}_{\alpha,N})$, then one can prove that $\mathcal{P}_{\alpha \beta,N+1} > \mathcal{P}_{\alpha \beta,N}$ by forming $(\mathcal{P}_{\alpha \beta,N+1} - \mathcal{P}_{\alpha \beta,N}) \cdot (\mathcal{P}_{\alpha,N+1} \cdot \mathcal{P}_{\alpha,N})$ and manipulating terms to show its positivity.}.

\begin{remark}
    Practical algorithms for evaluating logarithmic means typically reduce to an AEC-type scheme for small values of $\langle{\phi}\rangle$ \cite{Ismail_JCP_2009}. Without the solution-dependent adaptivity proposed herein, such procedures are technically indefinite in their entropy stability, although the consequences of this may be minimal due to the resulting violations being tied to very small jumps.
    \label{remark1}
\end{remark}

An entropy stable implementation of the AEC schemes is therefore possible via
\begin{equation}
\begin{array}{l l }
\meanless{\rho} =  \mean{\rho}^{<\log}_{H,AEC(N)} \ \left(\text{or} \ \  \mean{\rho}^{<\log}_{G,AEC(N)}\right), \qquad&
\meanmore{\rho} =  \mean{\rho}^{>\log}_{A,AEC(N)}, \\
\meanless{e} =  \mean{e}^{<H\log}_{H,AEC(N)}, &
\meanmore{e} =  \mean{e}^{>H\log}_{A,AEC(N)}  \ \left(\text{or}  \ \  \mean{e}^{>H\log}_{G,AEC(N)}\right)
\end{array} \ ,
\end{equation}
while employing the adaptivity guidelines of Table \ref{tab:ES_conditions}.
PEP is furthermore enforced due to the necessary requirements of the associated modifying functions from Eq.~\eqref{eq:PEP_requirements_modifying} being satisfied.
Assuming a $N$ level expansion, then Eq.~\eqref{eq:ES_conditions} suggests the entropy stabilization terms to take the form $\sim \sum_{n>N}|(\delta \rho)^{2n+1} |$ and $\sim\sum_{n>N}|(\delta 1/T)^{2n+1}|$.
This differs notably from standard ES mechanisms which induce even power gradients within the entropy balance law.

\section{Results}

In this section, two benchmark tests are used to evaluate whether the proposed method yields positive entropy production and preserves pressure equilibrium. Additionally, its potential in mitigating oscillations in the presence of shocks is examined.
Both tests will use 2-order accurate fluxes and high-order time integration with small time-stepping (e.g., a fourth-order Runge--Kutta time integrator with $\text{CFL} = 0.01$) to mitigate temporal discretization error effects.

\begin{figure}[tb]
    \centering
\begin{subfigure}[b]{0.33\textwidth}
    \centering
    \includegraphics[width=\textwidth]{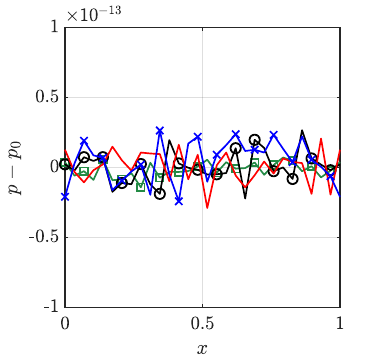}
    \caption{Variation of pressure at time $t=5$.}
    \label{fig:DW_p}
\end{subfigure}
\begin{subfigure}[b]{0.33\textwidth}
    \centering
    \includegraphics[width=\textwidth]{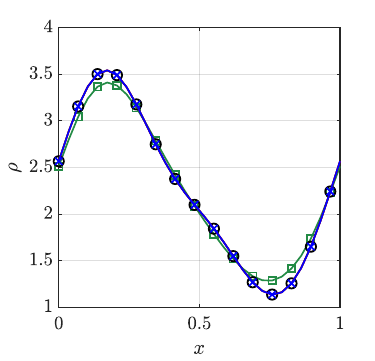}
    \caption{Density at time $t=5$.}
    \label{fig:DW_rho}
\end{subfigure}
\begin{subfigure}[b]{0.33\textwidth}
    \centering
    \includegraphics[width=\textwidth]{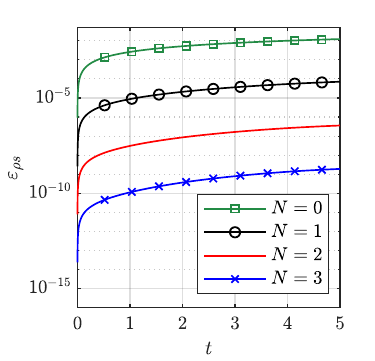}
    \caption{Time evolution of the global entropy.}
    \label{fig:DW_rhos}
\end{subfigure}
    \caption{Simulation of a density wave using the ES AEC schemes with a varying number of expansion terms $N$.}
    \label{fig:DW}
\end{figure}

\begin{figure}[tb]
    \centering
\begin{subfigure}[b]{0.24\textwidth}
    \centering
    \includegraphics[width=\textwidth]{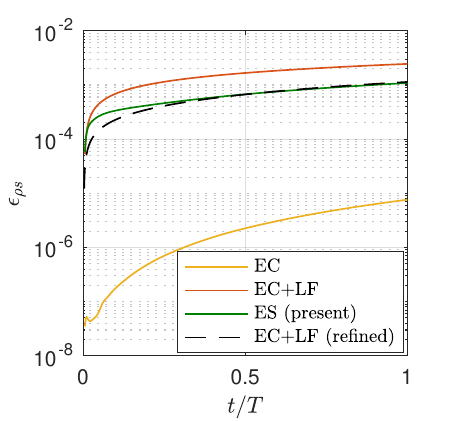}
    \caption{Global entropy evolution.}\label{fig:Sod_globent}
\end{subfigure}
\begin{subfigure}[b]{0.24\textwidth}
    \centering
    \includegraphics[width=\textwidth]{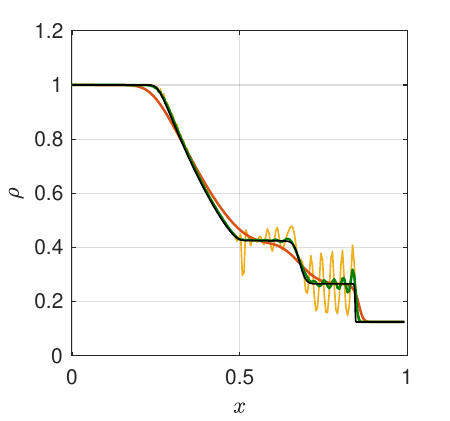}
    \caption{Density $(\rho)$}\label{fig:Sod_rho}
\end{subfigure}
\begin{subfigure}[b]{0.24\textwidth}
    \centering
    \includegraphics[width=\textwidth]{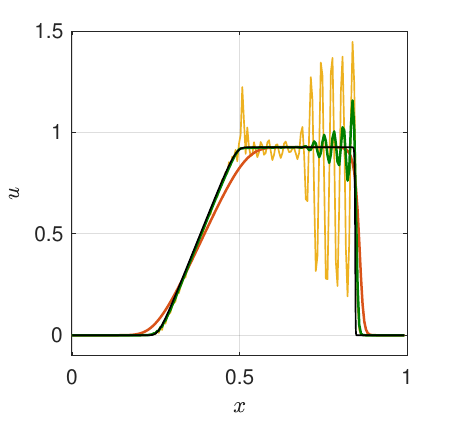}
    \caption{Velocity $(u)$}\label{fig:Sod_u}
\end{subfigure}
\begin{subfigure}[b]{0.24\textwidth}
    \centering
    \includegraphics[width=\textwidth]{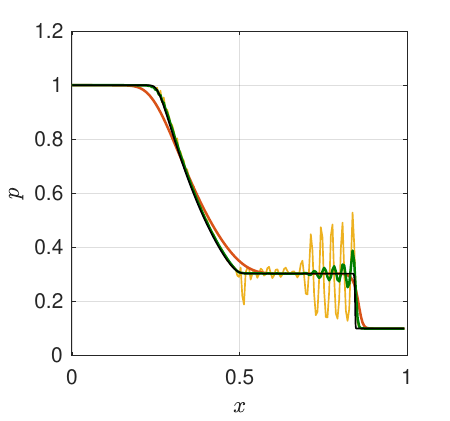}
    \caption{Pressure $(p)$}\label{fig:Sod_p}
\end{subfigure}
    \caption{Sod shock tube results, comparing baseline EC with EC-LF stabilization and the proposed adaptive averaging with $\mathcal{M}_\text{sc}$.
    }
    \label{fig:Sod}
\end{figure}

\paragraph{Density wave}
The first test is that of a density wave in a 1D periodic domain $\Omega=[0,1]$ with initial conditions as in \cite{DeMichele_JCP_2023}: $(\rho,u,p)_0 = \left(1 + \exp\left(\sin\left({2\pi x}\right)\right),1,1\right)$, meaning that the density wave loops with a period equal to 1.
We evaluate the ES version of the AEC schemes using a varying number of expansion terms, ranging from $N=0$ to $N=3$.
A key objective of this test is to verify the PEP property.
As shown in Fig.~\ref{fig:DW_p}, this property is upheld, in agreement with theoretical predictions.
For all variations of our proposed AEC scheme tested here, the pressure remains constant at its initial value, exhibiting only a machine-precision roundoff error on the order of \num{e-14} after five full loops. 
Examining the density profile at $t=5$ in Fig.~\ref{fig:DW_rho}, we observe that the wave's amplitude is slightly reduced for the $N=0$ formulation; on the other hand, the profiles produced by all other schemes ($N\geq 1$) are virtually indistinguishable. 
Finally, we assess the schemes' performance regarding entropy production, which is a primary focus of our proposed methodology.
As established in the theoretical framework, entropy production is expected to be strictly positive, and our numerical results confirm this behavior.
Figure~\ref{fig:DW_rhos} illustrates the time evolution of the relative global entropy variation with respect to the initial global value, defined as
 $\varepsilon_{\rho s} = \left(\int_\Omega \rho s \, \d \Omega\right)\left/\left(\int_\Omega \rho_0 s_0 \, \d \Omega\right)\right. - 1$. 
 This metric highlights how the number of expansion terms N can be used to calibrate the amount of entropy produced. Notably, at time $t=5$, the relative entropy variation decreases by several orders of magnitude, dropping from \num{e-2} for $N=0$ down to \num{e-9} for $N=3$.

\paragraph{Sod shock tube}
The second test case is the 1D Sod shock tube \cite{Sod_JCP_1978} with initial conditions $(\rho,u,p)_\text{left} = (1,0,1)$ for $x \le L/2$ and $(\rho,u,p)_\text{right} = (0.125,0,0.1)$ for  $x > L/2$, discretized with 200 uniform grid points across the domain length $L$.
In terms of the proposed procedure, We employ the density and internal energy averages as the modified arithmetic and harmonic means of Eq.~\eqref{eq:modifying_functions} with adaptation according to Table \ref{tab:ES_conditions}. 
In order to achieve mollification of sharp gradients, the modifying functions are designed to move the averages \emph{away} from the logarithmic means, thus increasing the entropy stabilization\footnote{As an alternative to achieving ES by modifying the simple means, one can modify the native EC logarithmic means directly (albeit with the caveat from Remark \ref{remark1}) as $\meanmore{\rho} = \logmean{\rho} \cdot \mathcal{M}$, $\meanless{\rho} = \logmean{\rho} / \mathcal{M}$, $\meanmore{e} = \mean{e}^{H\log} \cdot \mathcal{M}$, and $\meanless{e} = \mean{e}^{H\log} / \mathcal{M}$, with $\mathcal{M}(\phi_i,\phi_{i+1}) \ge 1$ . This structure also maintains PEP (i.e., $\meanone{\rho}\meantwo{\rho^{-1}} = 1)$.}. 
We consider the function $\mathcal{M}_\text{sc}(\phi) = \left[1 + 2 \cdot (\log r)^2 \right]$ with $r \triangleq (\max\{\phi_i,\phi_{i+1}\}/\min\{\phi_i,\phi_{i+1}\}) \ge 1$ for shock capturing, which increases penalization for larger jumps.
We note that other functions are possible but that no formal attempt has been made to optimize $\mathcal{M}_\text{sc}$, which may constitute a line of future research.
In order to further enforce PEP per Eq.~\eqref{eq:PEP_requirements_modifying}, we choose $\mathcal{M}_\text{sc} = \mathcal{M}_A^{> \log} = \mathcal{M}_A^{> H\log} = 1/\mathcal{M}_H^{< \log} = 1/\mathcal{M}_H^{< H\log}$.
The novel ES regularization is then compared to the EC flux of Ranocha as well as an ES rendition recovered by adding a standard local Lax--Friedrich (LF) dissipation \cite{Chandrashekar_CCP_2013}, which yields entropy stability for two-point fluxes but lacks further structures such as energy consistency and PEP. 
Also included is a LF solution calculated on a 2000 point grid, which is taken as a notional reference. 
Figure \ref{fig:Sod_globent} confirms the global entropy stability of the ES methods, with the nominal LF method showing greater regularization. 
Meanwhile Figs.~\ref{fig:Sod_rho}--\ref{fig:Sod_p} plot solutions at the final time $t = 0.2t_c$ and show how the new approach successfully reduces noise throughout the domain compared to the baseline EC Ranocha method, including capturing the shock at $x/L \sim 0.85$ and contact discontinuity at $x/L \sim 0.65$. 
Notably, the new method outperforms the LF dissipation through the expansion fan region ($x/L \in [0.25,0.45]$) as well at the contact discontinuity; impressively, it is comparable to the LF solution on the fine grid in these areas.
However the new procedure exhibits Gibbs oscillations near the shock, perhaps indicative of its dispersive mechanisms for entropy stability.
Future work can investigate improving such shock capturing performance through the careful design of the modifying function, $\mathcal{M}_\text{sc}$. 
%

\section{Conclusions}

A new procedure for recovering entropy stable discretizations for the compressible Euler equation has been developed.
By adapting the density and internal energy reconstructions based on local products of the velocity and key thermodynamic gradients, we recover provable point-wise stabilization to the entropy dynamics.
Additionally, the formulation naturally satisfies KEP and can be made to maintain PEP.
Herein, the strategy is used to yield an entropy-stable implementation of the AEC schemes as well as a novel strategy for shock regularization.
By employing central averages within the convective fluxes, the proposed methodology is distinctly different from traditional dissipation-based ES techniques.

Several refinements and extensions to the new schemes are of interest to explore within the context of compressible flows, including: accounting for overall stability purely through the density averaging, application of the fluxes as alternative ES boundary conditions and interface coupling for multi-element/block configurations \cite{Svard2014,Winters2026}, and further refinement of the shock-capturing modifying function $\mathcal{M}_\text{sc}$.
Extension to multi-component and non-ideal gases is also of interest and stems directly from the current development \cite{Aiello_JCP_2025,Aiello_TPG_2026,Klein2026}.
While the present note has focused on two-point fluxes, extension to high-order via flux-differencing \cite{Pirozzoli_JCP_2010,Fisher_JCP_2013b} is possible and would require one to account for the sign of the difference coefficients; this implies an asymmetric two-point flux, where the ES contribution can nevertheless be made to maintain the desired order provided that a sufficiently accurate expansion of the means is employed. 

\section*{Acknowledgments}
AE acknowledges funding for this work from the Air Force Office of Scientific Research (AFOSR) (program officers: Drs. Chiping Li, Fariba Farhoo, and Justin Koo) under contract No. 25RQCOR004, as well as Amentum under contract No. FA9300-20-F-9801.

\appendix

\bibliographystyle{model1-num-names}
\bibliography{Biblio_KEP_Compr}

\end{document}